\documentclass{article}

\usepackage[dvips]{graphicx}

\input{tcilatex}

\begin{document}

\title{\mbox{\Large The Equivalence Principle and the bending of
light}}

\author{\noindent Rafael Ferraro\thanks{
ferraro@iafe.uba.ar} \\
%EndAName
\\
{\small Instituto de Astronom\'\i a y F\'\i sica del Espacio,}\\
{\small Casilla de Correo 67, Sucursal 28, 1428 Buenos Aires,
Argentina}\\
\\
{\small Departamento de F\'\i sica, Facultad de Ciencias Exactas y
Naturales,}\\ {\small Universidad de Buenos Aires, Ciudad
Universitaria, Pabell\' on I,}\\ {\small 1428 Buenos Aires,
Argentina}}

\date{}

\maketitle

\begin{abstract}
The apparent discrepancy between the bending of light predicted by
the equivalence principle and its corresponding value in general
relativity is resolved by evaluating the deflection of light with
respect to a direction that is parallel transported along the ray
trajectory in 3-space. In this way the bending predicted by the
equivalence principle is fulfilled in general relativity and other
alternative metric theories of gravity.
\end{abstract}

\maketitle

\vskip1cm

\section{Introduction}\label{intro}
In 1911, five years before general relativity was formulated,
Einstein used the equivalence principle to calculate the bending
of a light ray in a weak gravitational field \cite{einstein}. By
considering the Doppler shift suffered by a light ray in a
reference system that moves with acceleration ${\bf a}=-{\bf g}$
in the absence of a gravitational field, Einstein deduced that the
frequency of light must experience a shift $\nu (0)=\nu (H)
(1+gH/c^{2})$ when the light moves a height $H$ in a weak static
gravitational field ${\bf g}$ ($gH/c^{2}<<1$). Because Einstein
did not yet possess a metric theory of gravity, he reasoned that
the gravitational redshift compels us to use ``clocks of unlike
constitution for measuring time at places with differing
gravitational potential'' \cite{einstein}. In fact, because the
number of wave maxima per unit of time should not vary at
different heights in a static gravitational field, a clock must go
$(1+gH/c^{2})$ times more slowly at $z=H$ than at $z=0$. Einstein
concluded that the velocity of light is $c$ when it is measured
with clocks of equal constitution; instead, the use of clocks of
unequal constitution implies that the velocity of propagation
should increase at the same rate as the clocks go slower:
$c(H)=c(0) (1+gH/c^{2})$. (``The principle of constancy of the
velocity of light holds good according to this theory in a
different form from that which usually underlies the ordinary
theory of relativity'' \cite{einstein}). Thus a plane wave
traveling transversely to a gravitational field changes its
direction of propagation -literally, it falls- because the
secondary fronts of the Huygens' construction advance faster where
the gravitational potential is higher. In fact, the deflection
$\phi $ in Fig.~1 is
\begin{equation}
\phi \simeq {\frac{[c_2-c_1] \Delta t}{H}}\simeq
{\frac{gH/c^{2}}{H} c} \Delta t\simeq gc^{-2}\Delta x\,.
\label{onda}
\end{equation}

Remarkably, the result in Eq.~(\ref{onda}) can also be obtained by
a direct application of the equivalence principle to the
trajectory of a particle traveling at the speed of light. In the
accelerated system (${\bf a}=-g\hat{ {\bf z}}$) such a particle
follows the equations of motion $ x\simeq ct$ and
$z=-\frac{1}{2}gt^{2}$. Therefore, its trajectory is $ z\simeq
-\frac{1}{2}gc^{-2}x^{2}$, so its slope after a displacement
$\Delta x$ is $dz/dx\simeq -gc^{-2} \Delta x$ in agreement with
Eq.~(\ref{onda}).

It is frequently emphasized that Eq.~(\ref{onda}) does not agree
with the experimentally tested bending of light predicted by
general relativity \cite{defle}, but only equals half this value
(see for instance Refs.~\cite{lanczos} and \cite{weinberg}). This
issue has deserved much attention, and it is usually said that the
equivalence principle contributes one half to the deflection while
the other half comes from the spatial curvature \cite{parker}. In
fact, the equivalence principle only prescribes the form
$g_{00}=1+2\Phi c^{-2}$ for the temporal part of the metric
associated with a weak gravitational potential $\Phi ({\bf r)}$,
but says nothing about the spatial geometry, which certainly
influences a light ray. However, this way of relating the
equivalence principle to the relativistic bending of light
diminishes the value of the equivalence principle by suggesting
that the equivalence principle is unable to predict a quantitative
physical result.

To reinstate the equivalence principle as a full statement about
the local behavior of matter and radiation in a gravitational
field so that it can yield results in quantitative agreement with
those from a metric theory of gravity, we should first establish
the kind of deflection that is referred to by the equivalence
principle in the context of a metric theory of gravity (see
Ref.~\cite{comer} for an approach of this type). In general, the
notion of deflection of a trajectory relative to a {\it fixed}
direction entails the parallel transport of that direction along
the trajectory. As it is well known, the angle between a
trajectory and a parallel transported direction does not change
when the trajectory is a geodesic. Instead, if the trajectory is
not a geodesic, then the angle will change (the trajectory {\it
deflects}). The world line of a ray light is a geodesic of the
4-dimensional spacetime. However, the equivalence principle for
the bending of light does not refer to the world line, but
contains a statement about the ray trajectory in 3-space. As we
will see, the trajectory in 3-space is not a geodesic. Thus the
angle between the trajectory and a parallel transported direction
changes along the trajectory. Therefore, the equivalence principle
should be regarded as a statement about this angle, which is a
geometric (invariant) object of 3-space.

Our first task will be to determinate the trajectory of a light
ray in the curved 3-space, together with its affine parameter
(that is, the length of the trajectory). Then we will calculate
the parallel transport of a vector along the trajectory of the
light. Finally we will show that a light ray deflects at the angle
predicted by the equivalence principle.

\section{The trajectory of a light ray}
We will work in a weak gravitational field geometry described by
the static metric
\begin{equation}
ds^{2}=(1+2\Phi({\bf r}) c^{-2})c^{2}dt^{2}-(1-2\gamma \Phi({\bf
r}) c^{-2})(dx^{2}+dy^{2}+dz^{2})\,, \label{metric}
\end{equation}
where the 3-space metric includes the post-newtonian parameter
$\gamma$ that is equal to 1 in general relativity \cite{mtw,will}.
We can use $\gamma$ to track the spatial geometry contribution to
the bending of light.

The behavior of a light ray is obtained from the equation for null
geodesics. Equivalently, we can start from the first integral
\begin{equation}
g^{ij} p_{i} p_{j}=0, \qquad p_{i}\equiv g_{ik}\frac{dx^{k}}{
d\lambda }, \label{momentos}
\end{equation}
where $\lambda $ is an affine parameter. Let us consider a ray
satisfying the initial conditions $dy/d\lambda = 0$ and
$dz/d\lambda = 0$. For simplicity, we will consider a region where
the gravitational field is nearly uniform: $\Phi =gz$, so $p_{0}$,
$p_{x}$, and $p_{y}$ are conserved magnitudes. Initially (at
$x=0$, $z=0$) Eq.~(\ref{momentos}) reads
\begin{equation}
\label{4} p_{0}^{2}-p_{x}^{2}=0\,.
\end{equation}
Because $p_{0}$ and $p_{x}$ are conserved, Eq.~(\ref{4}) implies
that $p_{0}^{2}=p_{x}^{2}$ for all value of $z$. Therefore,
Eq.~(\ref {momentos}) becomes
\begin{equation}
(1+2 gzc^{-2})^{-1} p_{x}^{2} - (1-2 \gamma gzc^{-2})^{-1}
p_{x}^{2} - (1-2 \gamma gzc^{-2}) \left( \frac{dz}{d\lambda
}\right)^{2} =0.
\end{equation}
If we retain only terms to first-order in $gzc^{-2}$, we have
\begin{equation}
-2(1+\gamma ) gzc^{-2} p_{x}^{2} \simeq \left( \frac{dz}{d\lambda
} \right)^{2},
\end{equation}
and
\begin{equation}
z=-\frac{1+\gamma }{2} p_{x}^{2} gc^{-2} \lambda^{2}+{\it
O}(\lambda^{4}). \label{z}
\end{equation}
Because
\begin{equation}
\frac{dx}{d\lambda }=g^{xx} p_x =-(1-2\gamma g z c^{-2})^{-1}
p_{x}\,,
\end{equation}
implies that
\begin{equation}
x = -p_{x} \lambda +{\it O} (\lambda^{3}),
\end{equation}
we have
\begin{equation}
z(x)=-\frac{1+\gamma }{2} gc^{-2} x^{2}+{\it O}(x^{4}).
\label{trayectoria}
\end{equation}
After a displacement $\Delta x$, the slope of the trajectory
$z(x)$ is
\begin{equation}
\frac{dz}{dx}\simeq -(1+\gamma ) gc^{-2} \Delta x.
\end{equation}
This result clearly separates the equivalence principle
contribution due to $g_{00}$ from the one associated with the
spatial geometry which is proportional to $\gamma$. Both
contributions are equal in general relativity.

\section{Bending of light in 3-space}
Any trajectory in the 3-space can be characterized by its tangent
3-vector $\bar{U}$ whose components are $\bar{U}^\alpha =
dx^\alpha /dl$ ($\alpha=1,2,3$); the affine parameter $l$ is the
length of the trajectory. In the 3-space of the static metric
(\ref{metric}), the infinitesimal length of the trajectory
(\ref{trayectoria}) is
\begin{equation}
dl^{2}=(1-2\gamma \Phi c^{-2})(dx^{2}+dz^{2})=(1-2 \gamma g c^{-2}
z(x)) [1+(dz/dx)^2] dx^{2}\,.
\end{equation}
Thus, $dl$ is
\begin{equation}
dl=[ 1+\frac{1}{2} (1+2\gamma ) (1+\gamma ) g^{2}c^{-4} x^{2}+
O(x^{4})] dx\,,
\end{equation}
and the 3-vector $\bar{U}$ that is tangent to the ray trajectory
is

\begin{eqnarray}
\bar{U}^{x} &=&\frac{dx}{dl}=1-\frac{1}{2} (1+2\gamma) (1+\gamma )
g^{2}c^{-4} x^{2}+O(x^{4})\nonumber \\ \bar{U}^{y} &=&
\frac{dy}{dl}=0 \\ \bar{U}^{z}
&=&\frac{dz}{dl}=\frac{dz}{dx}\frac{dx}{dl}=-(1+\gamma ) gc^{-2}
x+O(x^{3})\nonumber\,.
\end{eqnarray}

We can easily verify that the covariant derivative $D\bar{U}/Dl$
does not vanish, so the trajectory of the light ray is not a
geodesic in 3-space.

As mentioned in Sec.~\ref{intro}, the bending of light is measured
by the angle between the trajectory and a direction that is
parallel transported along it. So we are going to parallel
transport the 3-vector $\bar{V}=(\bar{V}^{x}, 0,\bar{ V}^{z})$
along the 3-curve described by the tangent vector $\bar{U}$. This
parallel transport is defined by the equation
\begin{equation}
\left(\frac{D\bar{V}}{Dl}\right)^{\alpha }\, \equiv\,
\frac{d\bar{V}^{\alpha }}{dl}+\bar{\Gamma }_{\beta \delta
}^{\alpha } \bar{U}^{\beta } \bar{V}^{\delta }=0\,,
\label{paralelo}
\end{equation}
where $\bar{\Gamma }_{\beta \delta }^{\alpha }$ are the
Christoffel symbols associated with the 3-space geometry. The only
non-zero Christoffel symbols in which we are interested are
\begin{equation}
\bar{\Gamma }_{xz}^{x}\!=\bar{\Gamma }_{zx}^{x}=-\gamma gc^{-2}
,\;\bar{ \Gamma }_{xx}^{z}=\gamma gc^{-2} ,\;\bar{\Gamma }
_{zz}^{z}=-\gamma gc^{-2}\,.
\end{equation}

Let us choose $\bar{V}(0)=\bar{U}(0)=(1,0, 0)$ as the initial
condition at $x=0,$ $z=0$. Thus the solution of
Eq.~(\ref{paralelo}) is

\begin{eqnarray}
\bar{V}^{x} &=&1-\frac{1}{2} (1+2\gamma) \gamma g^{2}c^{-4} x^{2}+
{\it O}(x^{4})\nonumber \\ \bar{V}^{z} &=&-\gamma gc^{-2} x+{\it
O}(x^{3})\,.
\end{eqnarray}

The angle between the directions of $\bar{V}$ and the trajectory
is
\begin{eqnarray}
\cos \phi (x)&=&\frac{\bar{U}\cdot \bar{V}}{\left| \bar{U}\right|
\left| \bar{V}\right|}=\frac{\bar g_{\alpha\beta}U^\alpha
V^\beta}{[\bar g_{\alpha\beta}U^\alpha U^\beta]^{1/2}} \nonumber
\\
{}&&=(1-2 \gamma g z(x) c^{-2})^{1/2} (\bar{U }^{x}
\bar{V}^{x}+\bar{U}^{z} \bar{V}^{z}) (\bar{U}^{x}
\bar{U}^{x}+\bar{U}^{z} \bar{U}^{z})^{-1/2}\,,
\end{eqnarray}
where we utilized the fact that $\left| \bar{V}\right| =1$ is
preserved by the parallel transport in 3-space. Therefore
\begin{equation}
\cos \phi (x)=1-\frac{1}{2} g^{2}c^{-4} x^{2}+{\it O}(x^{4}),
\end{equation}
so the deflection angle after a small displacement $\Delta x$ is
\begin{equation}
\phi \simeq gc^{-2} \Delta x\,,
\end{equation}
in agreement with the result (\ref{onda}) obtained by using the
equivalence principle.

\section{Conclusion}
The use of the equivalence principle can satisfactorily account
for the bending of light in the context of the weak gravitational
field metric (\ref{metric}). The result from the equivalence
principle corresponds to a physically meaningful quantity: it is
the angle formed by the trajectory of the light ray and a
direction that is parallel transported along that trajectory in
3-space. The post-newtonian parameter $\gamma$ does not enter the
deflection angle, although it obviously affects both the
trajectory and the parallel transport. In this way, any metric
theory of gravity that leads to the weak field behavior of
Eq.~(\ref{metric}) contains the bending of light predicted by the
equivalence principle.

\vskip1cm

{\bf Acknowledgments}

This work was supported by Universidad de Buenos Aires (Proy.\
X143) and Consejo Nacional de Investigaciones Cient\'{\i}ficas y
T\'{e}cnicas (Argentina).

\begin{figure}[p]
\centering
\includegraphics{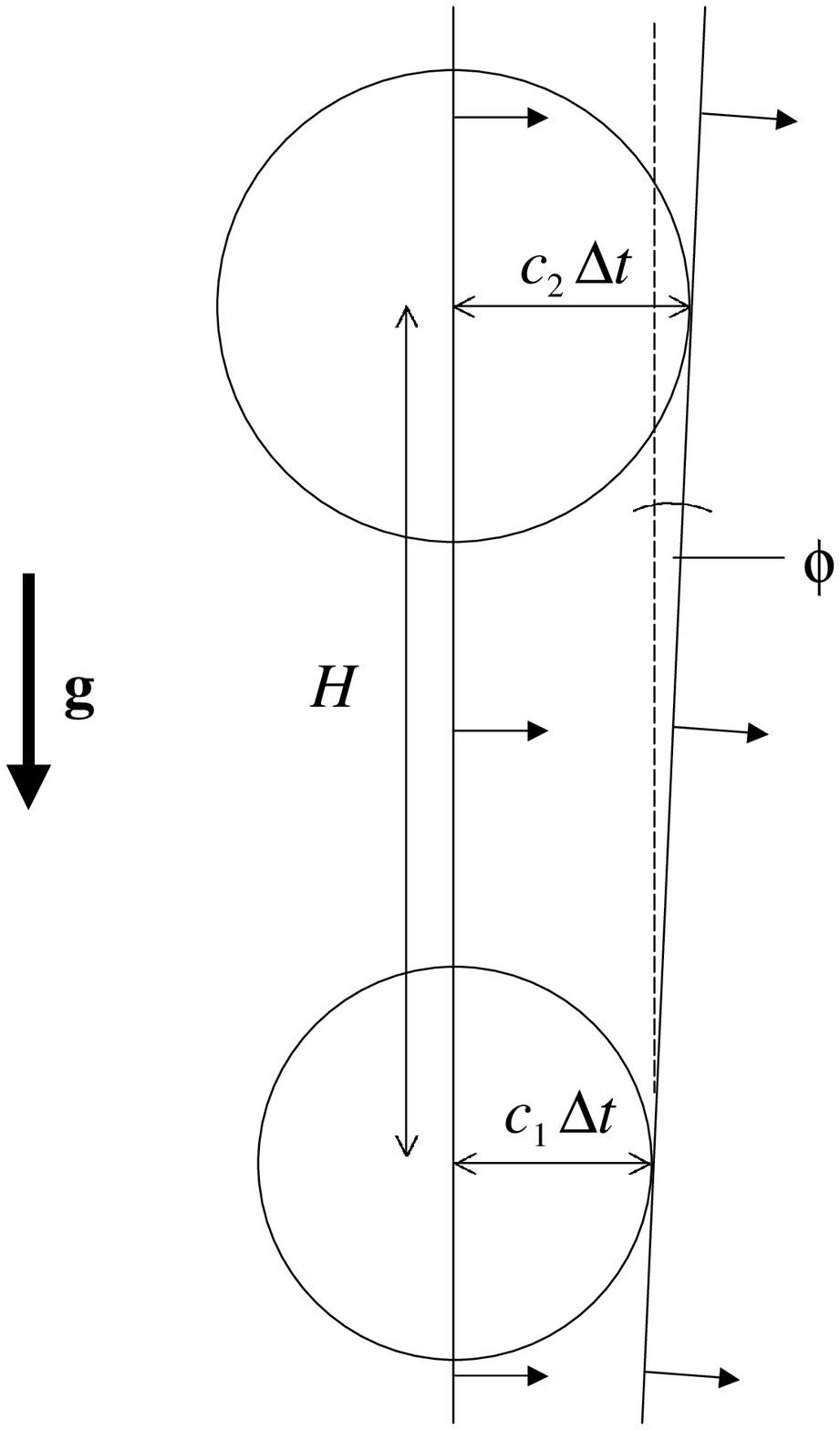}
\caption{Huygens construction of the wave front in a static
gravitational field.}
\end{figure}

\end{document}